\documentclass[pre,floatfix,twocolumn, 10pt]{revtex4-1} 
\usepackage{graphicx}
\usepackage{amssymb}
\usepackage{natbib}
\usepackage{dcolumn}
\usepackage{bm}
\usepackage{color}
\usepackage[colorlinks=true, linkcolor=blue, citecolor=blue]{hyperref}
\usepackage{subfigure}
\usepackage{graphicx,psfrag,xspace}
\usepackage{color}
\usepackage{amssymb,amsfonts,amsmath}
\usepackage{setspace}

\begin{document}

\author{E. A. Jagla} 
\affiliation{Comisi\'on Nacional de Energ\'{\i}a At\'omica, Instituto Balseiro (UNCu), and CONICET\\
Centro At\'omico Bariloche, (8400) Bariloche, Argentina}

\title{Different universality classes at the yielding transition of amorphous systems}

\begin{abstract} 

We study the yielding transition of a two dimensional amorphous system under shear by using a mesoscopic elasto-plastic model. The model combines a full (tensorial) description of the elastic interactions in the system, and the possibility of structural reaccommodations that are responsible for the plastic behavior. The possible structural reaccommodations are encoded in the form of a   ``plastic disorder" potential, which is chosen independently at each position of the sample to account for local heterogeneities.
We observe that the stress must exceed a critical value $\sigma_c$ in order for the system to yield.
In addition, when the system yields a flow curve relating stress $\sigma$ and strain rate $\dot\gamma$ 
of the form $\dot\gamma \sim(\sigma-\sigma_c)^\beta $ is obtained.
Remarkably, we observe the value of $\beta$ to depend on some details of the plastic disorder potential. 
For smooth potentials a value of $\beta\simeq 2.0$ is obtained, whereas for potentials obtained as a concatenation of smooth pieces a value $\beta\simeq 1.5$ is observed in the simulations. This indicates a dependence of critical behavior on details of the
plastic behavior that has not been pointed out before. In addition, by integrating out non-essential, harmonic degrees of freedom, we derive a simplified scalar version of the model that 
represents a collection of interacting Prandtl-Tomlinson particles.
A mean field treatment of this interaction reproduces the difference of $\beta$ exponents for the two classes of plastic disorder potentials, and provides values of $\beta$ that compare favorably with those found in the full simulations.

\end{abstract}

\maketitle

\section{Introduction}

Upon the application of a sufficiently large shear stress, any solid material will eventually yield. In the case of crystalline materials, yielding is produced by the motion of dislocation, which are defects 
of the otherwise perfect crystalline structure. In the case of amorphous materials, there is no such reference state on top of which imperfections can be easily defined. This has greatly delayed a theory of amorphous plasticity. 
However, as first recognized by Argon \cite{argon}, plasticity in this case can be defined in terms of discrete localized non-affine rearrangements that produce elastic stresses and can lead to a complex sequence of correlated deformations. These ideas have led to the development of the theory of shear transformations zones\cite{stz} that is nowadays one of the central concepts in amorphous plasticity. 

One of the hallmarks of amorphous plasticity is the existence of a yield point of the material, namely the existence of a minimum stress $\sigma_c$ that has to be exceeded in order to observe yielding.
In many cases, particularly for soft complex materials such as foams, pastes, etc., and also in the case of metallic glasses, it happens that for a fixed applied stress $\sigma$ beyond the yield point the material can reach a stationary condition of constant strain rate $\dot\gamma$. This allows to define the flow curve of the material $\dot\gamma(\sigma)$.
The nature of the yielding transition around $\sigma_c$ has been a matter of considerable interest. In the athermal case, in which the effect of thermal fluctuations is negligible, the most widely accepted view is that yielding corresponds to a well defined continuous transition at 
$\sigma_c$, such that $\dot\gamma=0$ for $\sigma<\sigma_c$, with $\dot\gamma$ increasing smoothly as $\sigma$ becomes larger than $\sigma_c$.
It is typically found \cite{exp1,exp2,exp3,exp4,exp5} that the dependence of $\dot\gamma$
near the yielding point has the Herschel-Bulkley form\cite{strain_loc} $\sigma-\sigma_c\sim \dot\gamma ^{1/\beta}$.  $\beta$ is known as the flow exponent and it is an important characteristic of the problem. 

An appealing idea to better understand the yielding transition has emerged from the comparison of this problem with the problem of depinning of elastic media moving onto disordered energy landscapes\cite{fisher,kardar}. In that case, the existence of a flow curve with a well defined $\beta$ exponent has been proven in a rather general way. One of the main conclusions of those studies is that the depinning transition corresponds to a critical point of the dynamics, at which the system becomes highly correlated and a diverging correlation length exists. This points out in particular to values of $\beta$ that are ``universal", depending in particular on the dimensionality $d$ of the system. For depinning $\beta\simeq 0.25$ in $d=1$ \cite{exponentes}, increasing for higher dimensions, and reaching the value $\beta=1$ in the mean field limit ($d\ge 4$).

A second similarity between depinning and yielding is in the form in which the dynamics proceeds close to the transition. In both cases an infinitesimal increase in the driving can produce an avalanche of activity. These avalanches are characterized  by its size and duration, and its distribution is an important characteristic of the problem.
Yet, an important difference between yielding and depinning is the following. While for depinning the advance of a small piece of the interface generates a positive effect on any other part of the system (trying to move forward the interface in any other point), for yielding the elastic interaction has effects of alternating signs in different parts of the sample. This fact (early considered by Eshelby\cite{eshelby}), has important consequences for the phenomenology of yielding, and is responsible for the existence of slip directions in which deformation can accumulate without producing any stress increase in the sample.

The formal analogy between the yielding problem and the depinning transition is thus an interesting line of investigation. Although there are clear numerical differences between the two cases (in particular, $\beta < 1$ for depinning, whereas 
$\beta> 1$ is systematically found for yielding), a scenario in which the yielding transition is supposed to correspond to a critical point with diverging correlation lengths has found much consensus\cite{pnas}, and triggered an important theoretical and experimental effort aimed at its verification.

Different numerical techniques have been applied to study the yielding transition, including direct atomistic simulations\cite{atom1,atom2,atom3,atom4,atom5}, and effective approaches such as soft glassy rheology\cite{sgr1,sgr2}, and elasto-plastic models\cite{ep1,ep2,ep3,ep4,ep5,ep6,ep7,ep8,ep9}. 
Elasto-plastic models are particularly suited to address the relation between yielding and depinning.
In these models the increase of plastic deformation in some region leads (through the action of a well defined elastic kernel)
to the modification of the elastic stress in other regions of the sample, which can produce new plastic re-arrangements.
In elasto-plastic models the long range elastic interaction is explicitly introduced
in the form of elastic propagators. Yet, the dynamical nature of the elastic interaction is
not fully accounted for and it is only effectively incorporated in the form of time delays for the interaction to propagate across the system. 

The model we are going to study shares many features with elasto-plastic models. In addition,
it incorporates in a more realistic way the elastic interactions through the system, and allows for a detailed description of the plastic deformation. Actually, one of the main findings will be that key properties of the model depend on the way in which plastic deformation evolves locally.
In particular, we find the value of the flow exponent $\beta$ to depend upon certain details of the disorder potential that is used to describe plasticity. Specifically, we find different $\beta$ values when the disorder potential has continuous second derivative ($\beta\simeq 2.0$, this case will be termed the ``smooth potential" case) and when it has points at which there are jumps of its first derivative ($\beta\simeq 1.5$, we call this case the ``parabolic potential" case). This unexpected non-unicity of the $\beta$ value is particularly important as it is obtained by changing a single characteristic of the model, and it cannot be related to artifacts originated in using different models, or different numerical techniques. This result challenges the idea of a single universality class of the yielding transition which, at least in this respect, seems to be less universal than its depinning counterpart. 

Trying to find a simple explanation of the results found, we transform the original model in an equivalent scalar problem that turns out to be a collection of interacting Prandtl-Tomlinson models\cite{p,t} (usually used to describe friction in elementary terms). 
By studying this model in different levels of approximation, we provide evidence that it
accounts for a yielding transition at a finite stress $\sigma_c$, and provides different $\beta$ exponents depending on the nature of the plastic potential used. Moreover, the actual values of $\beta$ found with the scalar model compare fairly well with those of the full tensorial simulations.

\section{Model}

The kind of modeling we are presenting originates in works of Bulatov and Argon\cite{ba}. It was generalized in different directions afterwards, and has been used to model a variety of non-linear problems of solids in which elasticity plays an important role. Examples include martensitic transformations\cite{martensite}, fracture patterns\cite{fracture} and elastic collapse of thin films\cite{elastic}. We have presented already the application of this technique to the modeling of yielding of plastic materials in \cite{jagla2007}, although in that case the focus was in the development of shear bands in the system when the material has some sort of structural relaxation. This last ingredient will not be incorporated here.

\begin{figure}
\includegraphics[width=7cm,clip=true]{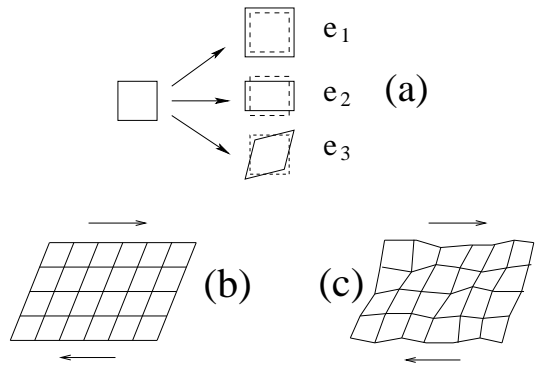}
\caption{(a) Definition of the three elementary distortions $e_1$, $e_2$, $e_3$ that describe the elastic state of the system at each spatial position. (b-c) Sketch of the state of a sample under an applied shear. (b) corresponds to the case of a system formed by identical elements, and (c) is the case in which each element has its own energy potential and energy minima.
\label{f0}
}
\end{figure}

We model a (two-dimensional) yielding plastic material as a collection of cells, each of them encoding the behavior of a large number of atoms or molecules in the system. The state of the cell is defined by its strain tensor $\epsilon_{ij}$.
It turns out to be more convenient to describe the elastic deformations by three independent strains
$e_1(r)$, $e_2(r)$, $e_3(r)$, representing volume distortions ($e_1$) and the two independent deviatoric distortions ($e_2$ and $e_3$) in the system (see Fig. \ref{f0}).
Values of $e_1$, $e_2$, and $e_3$ in different parts of the system are not independent. They satisfy a differential equation (known as the St Venant condition) that reads\cite{chandra}

\begin{equation}
(\partial^2_x+\partial^2_y)e_1-(\partial^2_x-\partial^2_y)e_2-2\partial_x\partial_ye_3=0
\label{compat}
\end{equation}

\begin{figure}
\includegraphics[width=7cm,clip=true]{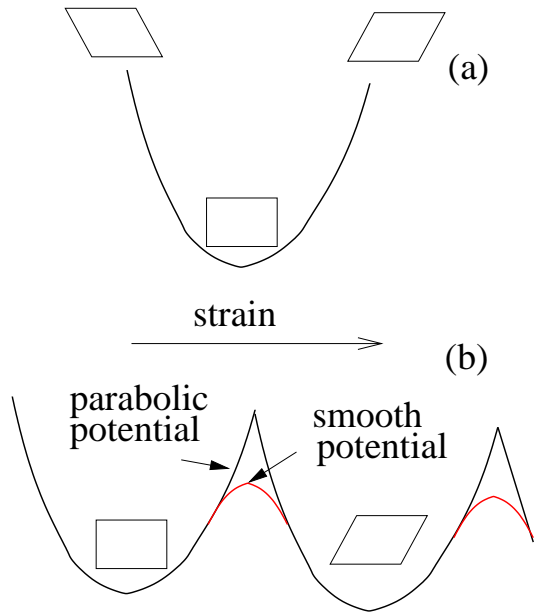}
\caption{Sketch of the local free energy depending on the strain state of a cell. (a) Perfectly elastic case. (b) Plastic case. In this case, other minima appear as the strain is increased further.
\label{f1}
}
\end{figure}

In order to describe the dynamics of the system it is necessary to define a free energy that depends on the strain state of all the cells.
If the system was a perfectly elastic, isotropic material, we would write a total free energy in the form
\begin{equation}
F=\int d^2r (Be_1^2+2\mu (e_2^2+e_3^2))
\label{uno}
\end{equation}
with $B$ and $\mu$ being the bulk and shear modulus of the material. 
However, to allow for the possibility to describe plastic deformation, the form of the free energy has to be modified. 
Referring to the sketches in Fig. \ref{f1} the free energy of a cell will increase upon deformation in the elastic regime (a), but eventually, it will reach a point in which a structural rearrangement occurs, and the free energy is reduced again to a new local minimum (b). It is assumed that 
structural rearrangements can continue to occur in a given cell when strain increases further, the local free energy thus consisting of a sort of 
``plastic potential", with different minima located at different values of deformation. The form of the potential near each minimum is quadratic, representing a local elastic state of the cell. For the transition between different local minima, we can consider at least two possibilities (see Fig. \ref{f1}). If we think of this  transition as some sort of irreversible rearrangement within the cell, a potential $V(e)$ consisting of a collection of parabolic pieces seems to be appropriate. This case will be called ``parabolic potential" case. However, we can consider also the case in which the first potential minimum gradually softens and eventually transforms smoothly into the next minimum. This is the case of a ``smooth potential".  One of the main findings of this paper is that the properties of the model depend crucially on the potential being ``smooth" or ``parabolic".

The strain values corresponding to the minima of the plastic potential are assumed to have stochastic values, which are different in different positions of the sample, leading to an interplay between elasticity and plastic disorder (sketched in Fig. \ref{f0}(c)) that is crucial for the behavior of the model.
We consider the model to be externally driven by applying a global deformation in one of the two deviatoric modes (we take it to be $e_2$, for concreteness\cite{e2e3}). For simplicity, we assume that plastic deformation in the system can appear only in the corresponding mode. This means that the quadratic part on $e_2$ of the free energy of an elastic solid (see Eq. \ref{uno}) will be replaced by an expression $V(e_2)$ describing the function in Fig. \ref{f1}(b), in such a way that the free energy is written as

\begin{equation}
F=\int d^2r(Be_1^2+2\mu e_3^2+V(e_2))
\label{dos}
\end{equation}
Details on how the functions $V(e_2)$ are actually constructed for the smooth and parabolic cases are given in an Appendix.
We only notice here that in order to preserve the isotropy of the model in the elastic limit, the form of $V(e_2)$ around any energy minimum is of the form $V(e_2)=2\mu (e_2-e_2^{min})^2$.

The dynamical evolution of the strains will be assumed to be overdamped. This will be reasonable
for sufficiently slow external variations of the control parameters, particularly the strain rate.
To be concrete, defining the local principal stresses $\sigma_i$ as
\begin{equation}
\sigma_i(x,y)=\frac{\delta F}{\delta e_i(x,y)},
\label{sigma-def}
\end{equation}
the dynamical evolution of the strain is 
obtained through a first order temporal evolution equation of the form
\begin{equation}
\frac{\partial e_i(x,y)}{\partial t}=-\varepsilon\sigma_i(x,y)+\Lambda_i(x,y,e_i,t)
\label{eqe}
\end{equation}
where $\Lambda_i$ is a Lagrange multiplier chosen to enforce the compatibility condition (\ref{compat}) \cite{martensite,jagla2007},
and $\varepsilon$ is the damping coefficient. In equilibrium ($\partial e_i(x,y)/\partial t=0$), 
this equation reduces to the standard elastic 
equilibrium equations, namely $\partial/\partial x_i~ (\delta F/\delta \epsilon_{ij})=0$ \cite{martensite}.




The numerical simulations presented here were performed under a constant externally applied rate of change of $\overline {e_2}$, namely $\overline {e_2}=\dot\gamma t$, and the main interest is in the evaluation of the corresponding stress $\sigma_2$. This is obtained from (\ref{sigma-def}) and (\ref{dos}), as ($\sigma_2$ will be simply noted $\sigma$, for simplicity):

\begin{equation}
\sigma =\overline{\frac{\partial V}{\partial e_2}}+\frac{\dot\gamma}{\varepsilon}
\label{sigma}
\end{equation}
where the bar indicates average over the sample, and the last term originates in the externally imposed zero-mode.
We scale $\sigma$ and $\dot\gamma$ in order to make $\varepsilon\equiv 1$, and also $\mu\equiv 1$ in Eq. (\ref{dos}).

\section{Results}

\begin{figure}
\includegraphics[width=8cm,clip=true]{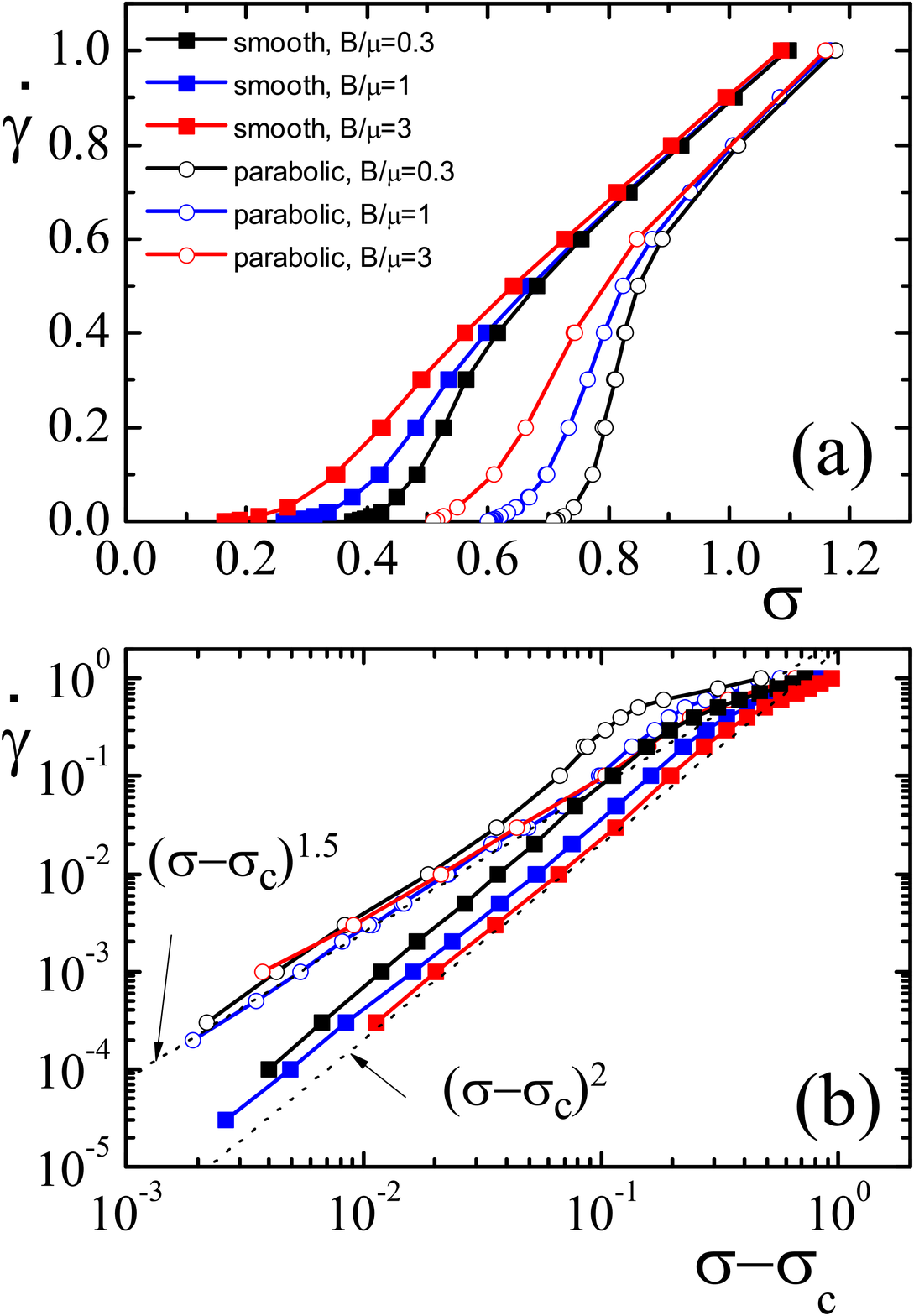}
\caption{Strain rate vs. stress curves, for systems with different values of $B/\mu$, for smooth and parabolic potentials. System size is 256$\times$256. (a) Linear scale. (b) Logarithmic scale with the value of $\sigma_c$ subtracted. Dotted lines are drawn for reference.
\label{f3}
}
\end{figure}

\begin{figure}
\includegraphics[width=8cm,clip=true]{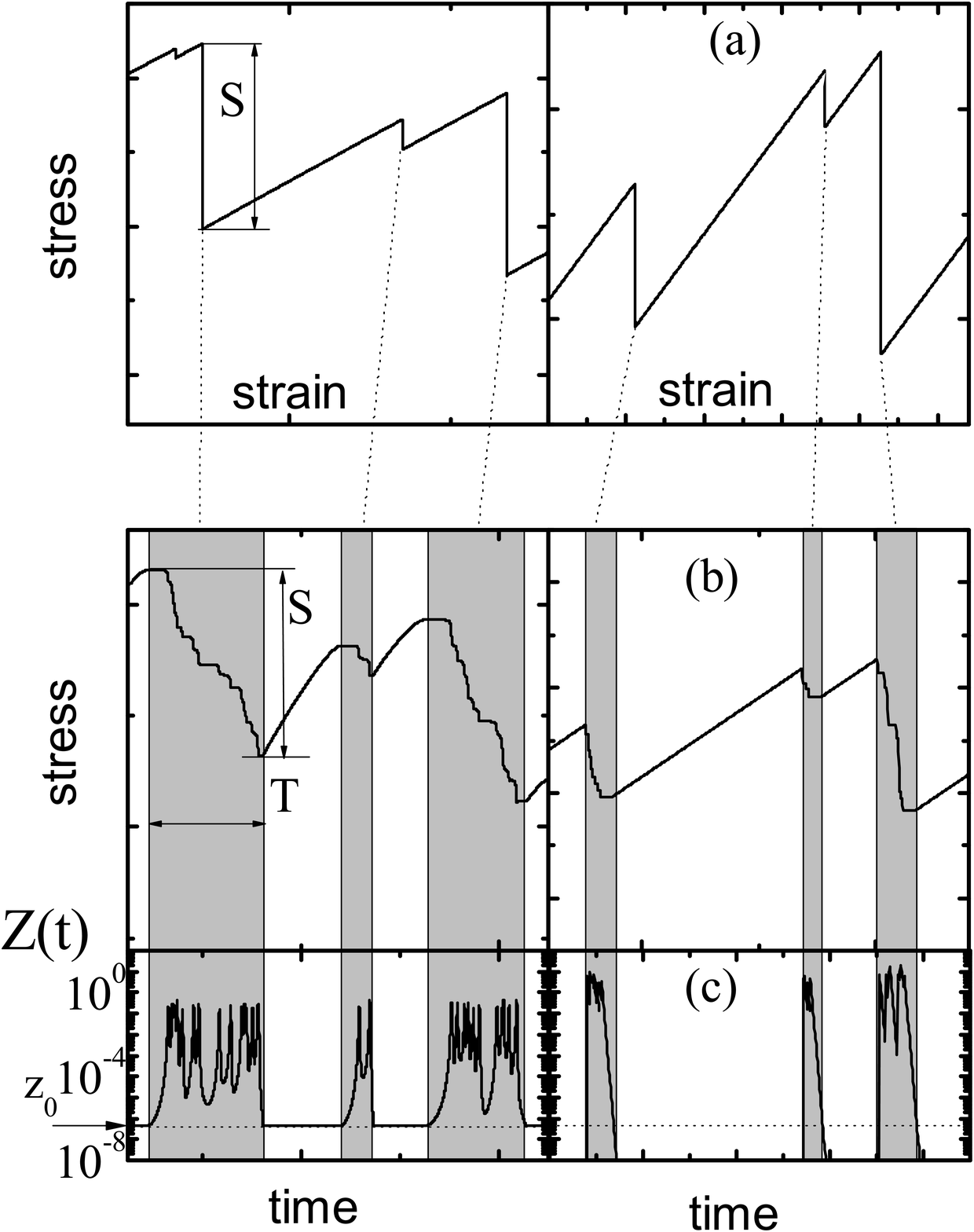}
\caption{Examples of the evolution of stress in the system, under the quasi-static protocol described in the text. Left part corresponds to smooth potentials, and right part to parabolic potentials. In (a) we see the stress-strain plot, and in (b) the stress-time one. Strain rate is zero in the gray regions (when $Z$, shown in panel (c), is larger than a threshold value $z_0$), whereas it is a fixed, small $\dot\gamma$ outside these periods. $T$ and $S$ measure the duration and size of the avalanches.
\label{sigma-t}
}
\end{figure}

In Fig. \ref{f3} we see the main results for the average stress in the system $\sigma$ as a function of the applied strain rate $\dot\gamma$. Results are presented for systems with different values of $B/\mu$, for smooth and parabolic potentials. The simulations clearly show the existence of a finite value $\sigma_c$ to which 
the stress converges as $\dot\gamma\to 0$, indicating the existence of a yield point in the model.
We observe that increasing $B/\mu$ systematically reduces the value of $\sigma_c$. In addition, we fitted  the lowest part of the curves ($\dot\gamma\leq 0.01$) with a form $\dot\gamma= C(\sigma-\sigma_c)^{\beta}$, adjusting $\sigma_c$, $\beta$ and $C$ to get the best fitting. 
The fitted values of $\beta$ for increasing values of $B/\mu$ are 1.61, 1.59, 1.43 for parabolic potentials, and 2.04, 1.92, 1.96 for smooth potential. Taking into account the numerical uncertainties, the conclusion is that the value of $\beta$ is independent of $B/\mu$, but it depends on the fact of using smooth or parabolic potentials. Although it is tempting to assign simple rational numbers to the values found (namely, $\beta=3/2$ for parabolic, and $\beta=2$ for smooth potentials), we stress that there is no reason, at the moment, to expect this is the case. 

Other quantities that are studied in models of the yielding transition have to do with the properties of individual avalanches close to the yielding point, when driving the system quasistatically. If driving is infinitely slow, the dynamics proceeds by a sequence of avalanches that are well separated in time, and that can be quantified by its size $S$ (which is defined as the stress drop in the system caused by the avalanche, see Fig. \ref{sigma-t}) and its duration $T$. In order to calculate these quantities in our model, and see in particular if they depend on the kind of potential used, we run quasistatic simulations in the following way. In a simulation with a small $\dot\gamma$, a quantity $Z$ measuring the rate of time evolution in the system is calculated.
We choose the quantity $Z$ to be $Z\equiv \sum {({\dot e_2})^2} $, where the sum runs over all sites of the system. $Z$ is very small when the system is in quasistatic equilibrium. However, when an avalanche is being triggered $Z$ rapidly increases. When this happens (in concrete, when  $Z$ exceeds some threshold value $z_0$) we stop the driving and follow the internal dynamics of the avalanche until $Z<z_0$ again. At this point driving is resumed until the next avalanche is triggered. 
In this way, we obtain stress-strain and stress-time curves as those shown in Fig. \ref{sigma-t}(a-b). Panel (c) shows the evolution of the quantity $Z$. It has to be noticed the difference in temporal evolution of $Z$ for the two kinds of potentials. In the parabolic case $Z$ has an abrupt jump up when a site goes over a cusp of the potential, initiating an avalanche. The avalanche ends with an exponential time decrease of $Z$. For the smooth potential case the evolution is much smoother. In particular, the beginning of an avalanche is marked by a progressive acceleration of $Z$ as one site passes over the smooth potential barrier. The finish of the avalanche is also more gradual in this case.

\begin{figure}
\includegraphics[width=8cm,clip=true]{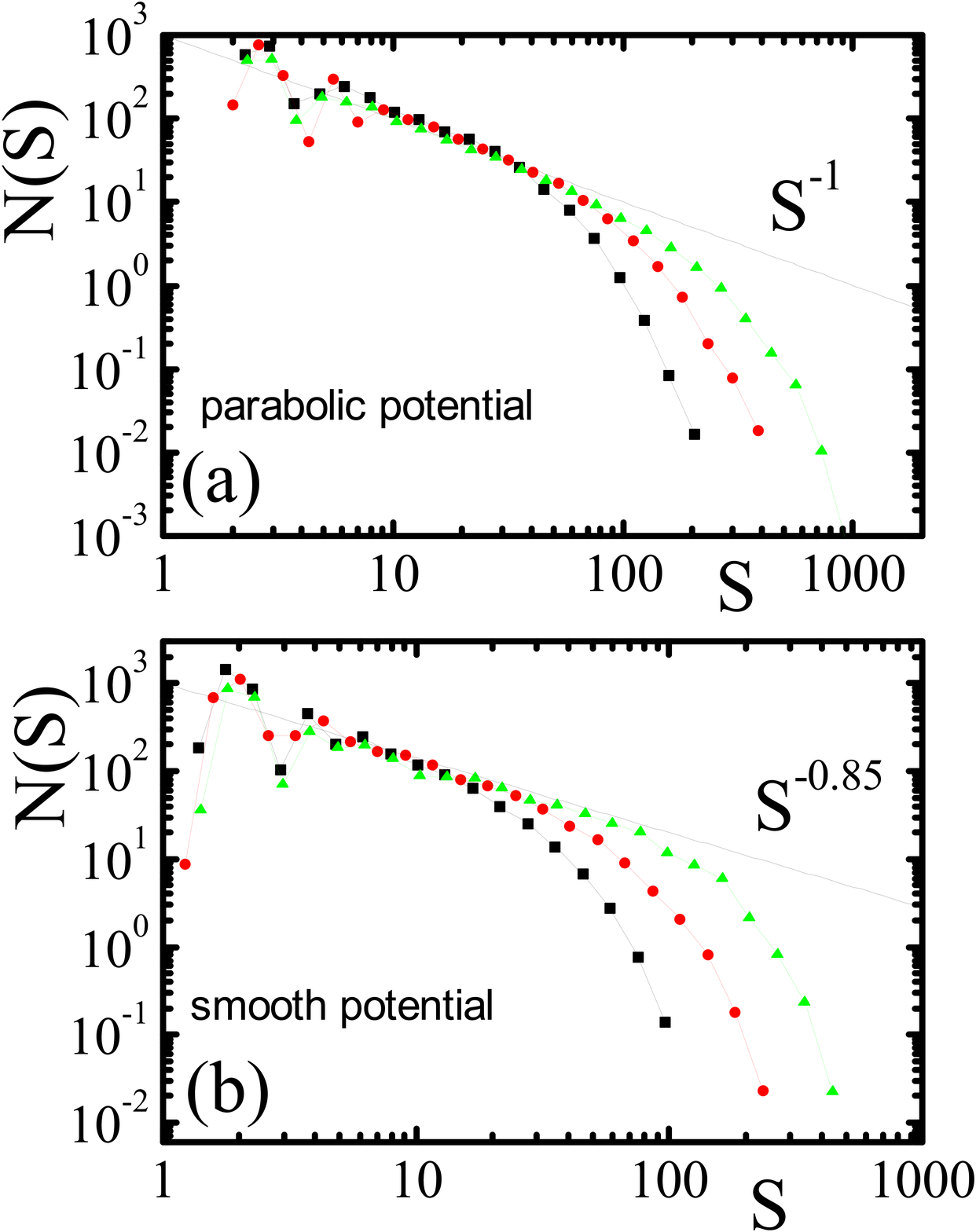}
\caption{Histogram of avalanche size distribution, in systems of different sizes, for (a) parabolic and (b) smooth potentials. The straight lines show some reference slopes.
\label{p-de-s}
}
\end{figure}

From curves as those in Fig. \ref{sigma-t}, a collection of avalanche sizes $S_i$, and avalanche durations $T_i$ can be obtained. These data are conveniently displayed in the following form. First of all we plot the histogram of avalanche size distribution in Fig. \ref{p-de-s}, where results for different system sizes are presented (from now on, all results presented correspond to $B/\mu=1$). We observe that the distribution is compatible with a power law distribution of avalanches $P(S)\simeq S^{-\tau}$, that is cut off at large avalanche sizes by the system size. The value of the exponent  $\tau$ is difficult to assess due to the small system sizes that we have been able to simulate. The reference power laws drawn in Fig. \ref{p-de-s} have lower slopes than values typically reported in the literature for the exponent $\tau$ (see a list of values in Table 2 of Ref. \cite{pnas}). We expect that simulations using larger system sizes will provide larger values of $\tau$.

\begin{figure}
\includegraphics[width=8cm,clip=true]{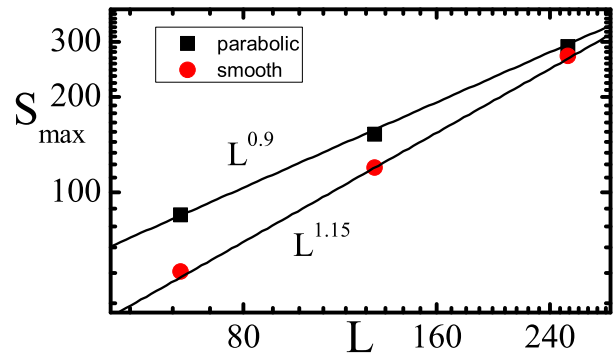}
\caption{The cut off avalanche size $S_{max}$ as a function of system size, for the smooth and parabolic potential cases. A dependence close to $S_{max}\sim L$ is observed in both cases.
\label{smax}
}
\end{figure}

On general grounds the scaling of the cutoff $S_{max}$ with the system size $L$ in the avalanche size distribution can be related to the fractal dimension ${d_f}$ of the avalanches. From the results in Fig. \ref{p-de-s} we can extract the value of $S_{max}$ as a function of $L$. The results are plotted in Fig. \ref{smax}. We observe that $S_{max}\sim L^{d_f}$ with ${d_f}$ slightly smaller than one for the 
parabolic potential ($d_f\simeq 0.9$), and slightly larger than one for the smooth potential ($d_f\simeq 1.15$). 
These results are compatible with values found in the literature \cite{pnas,salrob,saizer} (although larger values have also been reported \cite{nicolas,ep4}) and are naturally interpreted as originated in the fact that avalanches are correlated slip events along easy directions in the system,  which justifies its almost linear scaling with $L$.

\begin{figure}
\includegraphics[width=8cm,clip=true]{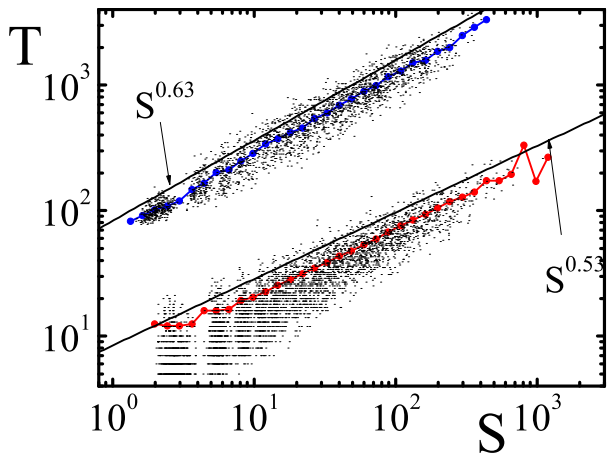}
\caption{Avalanche duration vs. avalanche size, for both kinds of potential, in a system of 256$\times$256.
Black dots are the results of individual avalanches, red (parabolic) and blue (smooth) curves are the average of $T$ in successive $S$ slices. Black lines are shown to display the overall behavior.
\label{z}
}
\end{figure}

A third result that can be obtained from curves such as those in Fig. \ref{sigma-t}, is the scaling between avalanche sizes and avalanche duration. This is plotted in Fig. \ref{z}. We see that $T_i$ vs $S_i$ shows a power law behavior $T_i\sim S_i^{p}$, with an exponent that differs slightly for both kind of potentials: $p\simeq 0.63$ for smooth potentials and  $p\simeq 0.53$ for the parabolic potential.
According to \cite{pnas} this exponent is $p=z/d_f$, and taking into account the previously found value of $d_f$, we obtain the values of the dynamical exponent as $z\simeq 0.75$ for smooth potentials and $z\simeq 0.5$ for parabolic potentials. We believe this difference between the two kinds of potentials is significant.

As a conclusion for this part, within the present accuracy of the simulations we are not able to tell if exponents $\tau$ and $d_f$ are different or not between the two kinds of potentials. However, the results for $z$ are more convincing, pointing to a difference between the two cases, in addition to the definitely different values of $\beta$ that we have found previously.

It is interesting to explore in the model some of the consequences of the alternating sign nature of the interaction kernel in the yielding problem (the Eshelby propagator\cite{eshelby}).  This is most easily seen in a single shear geometry: under the application of an external single shear, the deformation in the system does not need to be uniformly distributed. Actually, it can be localized in the form of a slip in a very narrow region of the system. Under some circumstances (requiring for instance some kind of aging of the material, see \cite{jagla2007}), the position at which deformation occurs can be persistent in time upon further application of the external stress, and a shear band in the system can be formed. However in the present case successive external deformation can be accommodated in the system in the form of slip between adjacent planes at different spatial locations\cite{nota_slip}. If these locations are uncorrelated in time, it can be expected that the strain increase in a given position of the system has the characteristics of a stochastic Poisson process. 
This analysis is also valid for the case in which the external deformation is a deviatoric stress, as in the present simulations, the only difference is that now deformation accumulates in a system of two different perpendicular slip directions (the $\pm 45 ^\circ$ directions in Fig. \ref{f0} when deformation is of the $e_2$ type).

\begin{figure}
\includegraphics[width=8cm,clip=true]{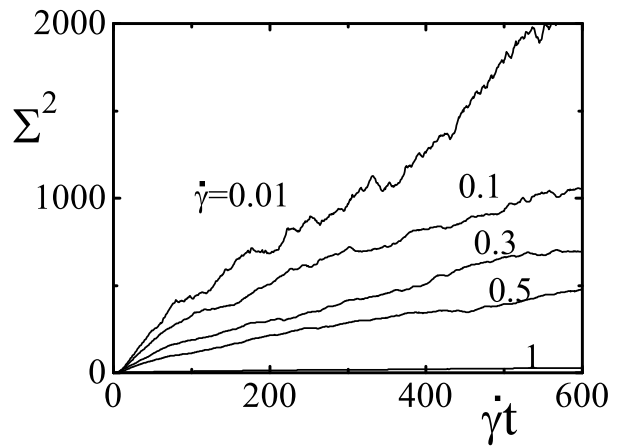}
\caption{Evolution of the variance of the strain in the system ($\Sigma^2=\overline{e_2^2}-\overline{e_2}^2$) as a function of the average strain $\overline{e_2}=\dot\gamma t$. Different curves were obtained for different values of $\dot\gamma$, as indicated. System size is $64\times 64$. 
\label{edet}
}
\end{figure}

In Fig. \ref{edet} we observe the evolution of the variance $\Sigma^2$ of the strain in the system as a function of the average strain itself. We see in fact how this quantity does not saturate but increases rather linearly with the applied total deformation. Note that the increase is more rapid when the value of $\dot\gamma$ is reduced. However, the results point clearly to an asymptotic maximum increase rate as $\dot\gamma\to 0$, indicating the existence of a quasi-static limit in which the external applied deformation is accommodated in an uncorrelated way in the system, leading to a typical diffusive increase of the strain fluctuation.

\section{Scalar description, and mean field analysis}

The finding of different critical exponents depending on details of the disorder potential is an unexpected result that deserves further analysis. The difference is definitely more clear in the case of the flow exponent $\beta$, where the numerical uncertainty of the results is smaller, and we concentrate on this in the following discussion.

We have been able to obtain an alternative, scalar description of the model that clarifies the origin 
of two different flow exponents for the two kinds of disorder potentials considered. In order to derive this alternative description  we reproduce here the basic equations of the model for clarity:

\begin{equation}
F=\int d^2r(Be_1^2+2\mu e_3^2+V(e_2))
\label{dos2}
\end{equation}

\begin{equation}
\dot e_i=-\varepsilon_i \frac {\delta F}{\delta e_i}+\Lambda_i
\label{tres2}
\end{equation}

\begin{equation}
(\partial^2_x+\partial^2_y)e_1-(\partial^2_x-\partial^2_y)e_2-2\partial_x\partial_ye_3=0
\label{compat2}
\end{equation}
(note that the damping coefficient $\varepsilon$ has been allowed to depend on the mode being considered).
An equivalent scalar description can be obtained by integrating out the harmonic degrees of freedom $e_1$ and $e_3$ in the previous equations. This can be easily done in the case in which the variables $e_1$ and $e_3$ equilibrate very rapidly compared to $e_2$ (i.e., $\varepsilon_1$, $\varepsilon_3\gg \varepsilon_2$), and this is the case that will be addressed here. This allows to search for the values of $e_1$ and $e_3$ that minimize the free energy, under the constraint given by Eq. (\ref{compat2}). A simple calculation in Fourier space shows that in this situation

\begin{equation}
B|e_{1{\bf q}}|^2+2\mu|e_{3{\bf q}}|^2= \frac{\mu B(q_x^2-q_y^2)^2}{\mu q^4+2B q_x^2q_y^2}|e_{2{\bf q}}|^2
\end{equation}
for any ${\bf q}\ne 0$. Now the model can be written as a single, unconstrained equation for $e_2$, which
in Fourier space  reads (${\bf q}\ne 0$)
\begin{equation}
\dot e_{2{\bf q}}=-\varepsilon_2 \left .\frac {dV}{d e_2}\right |_{\bf q}-\varepsilon_2G({\bf q})e_{2{\bf q}}
\end{equation}
with 
\begin{equation}
G({\bf q})=\frac{2\mu B(q_x^2-q_y^2)^2}{\mu q^4+2B q_x^2q_y^2}
\end{equation}
In order to write the model equation in real space, it is convenient to separate the average value of $G$ from its angular oscillating part. This leads to (we set $\varepsilon_2=1$)
\begin{equation}
\dot e_{2r}=f_r(e_{2r})+\sigma+k(\dot\gamma t -e_{2r})+\sum_{r'}\widetilde G(r-r')e_{2r'}
\label{tomlinson}
\end{equation}
($f_r(e_{2r})=-{d V_r}/{d e_{2r}}$), where 
\begin{equation}
k=\frac{1}{2\pi}\int_0^{2\pi}\frac{2\mu B(\cos^2(\theta)-\sin^2(\theta))^2}{\mu +2B \cos^2(\theta)\sin^2(\theta)}d\theta
\end{equation}
\begin{equation}
\widetilde G(r)=G(r)-k
\end{equation}
and the value of $\sigma$ 
is chosen as

\begin{equation}
\sigma=-\overline {f_r(e_{2r})}+{\dot\gamma}
\end{equation}
in order to satisfy the global constraint $\overline {e_2}=\dot\gamma t$.

The kernel $\widetilde G(r)$ has a $r^{-2}$ decay with distance, and a quadrupolar angular symmetry. It is noting but the Eshelby elastic propagator \cite{eshelby} producing a long range effective interaction in the $e_2$ field, mediated by $e_1$ and $e_3$. We emphasize however the appearance of the ``mean field like" term $k$ which couples all sites to the mean value of the strain in the system.
Note that same mean field like coupling has been obtained in the case of plasticity already in \cite{saizer}, and also
in other cases in which $e_2$ and $e_3$ are eliminated in favor of $e_1$ \cite{nelson,pepe}.

Eq. (\ref{tomlinson}) is very suggestive. In the absence of the last term, $e_{2r}$ is driven on top of the potential $V_r(e_{2r})$ by a spring of constant $k$. This is just the Prandtl-Tomlinson (PT) model used to qualitatively describe the origin of a friction force between sliding solid bodies\cite{p,t}. The main results that are obtained from the PT model in the absence of thermal fluctuations is the existence of a critical stress $\sigma_c$ for $\dot\gamma\to 0$ (as long as there are points at which $d^2 {V(e_2)}/{d e_2^2}>k$ ), and a power law increase of $\sigma$ for finite $\dot\gamma$, i.e, $\dot\gamma\sim (\sigma-\sigma_c)  ^\beta$. The value of $\beta$ turns out to be dependent of the kind of potential that is used. \cite{f2} 
For smooth potentials $\beta=3/2$, whereas for parabolic potentials (with points at which the first derivative has jumps) the value $\beta=1$ is obtained. 

In the presence of the last term, Eq. (\ref{tomlinson}) defines a set of coupled PT models, in which the variable $e_{2r}$ is driven by the external driving and by the effect of all the $e_{2r'}$ through the coupling term $\widetilde G(r-r')$. 
We are currently conducting simulations of Eq. (\ref{tomlinson}) in order to re-obtain within this framework the kind of results presented in Section III.
For the time being, in order to provide a mean-field-like approach to Eq. (\ref{tomlinson}) (see also \cite{wyart,lemaitre}), we will 
replace the distance-dependent coupling $\widetilde G(r-r')$ by a term that is only dependent on $r'$, i.e, the fluctuating term is supposed to be unique for all sites in the system.
Then we write the mean field equations in the form (we drop the subindex 2, for simplicity)

\begin{eqnarray}
\dot e_{\alpha}&=&f_{\alpha}(e_{\alpha}) +\sigma +k(\dot\gamma t-e_{\alpha})+w(t) \label{sc3}\\
w(t)&=&\sum_{\alpha} \lambda_{\alpha} e_{\alpha} \label{sc4}
\end{eqnarray}
where ${\alpha}=1,...,N$ labels the $N$ sites in the system, 
and the variables $\lambda_{\alpha}$ (with $\sum_\alpha \lambda_\alpha=0$) define how the self consistent driving term $w(t)$ is constructed in a unique way for the whole system\cite{hl}. 
In the limit of $N\to\infty$, the precise distribution of $\lambda_\alpha$ values in Eq. (\ref{sc4}) becomes irrelevant, and the values of $\lambda_\alpha$ can be taken from a normal distribution\cite{nota_lambda}. To ensure a correct thermodynamic limit we must choose
$\langle\lambda_\alpha^2\rangle\sim 1/N$.



Before analyzing this mean field model form for particular distributions of the variables $\lambda_{\alpha}$, we want to consider a simplified version of it for which we have found analytical expressions for the flow exponent $\beta$. This version is obtained by breaking the self-consistency condition, and taking the value of $w(t)$ in Eq. (\ref{sc4}) to be externally prescribed. In order to define the statistical properties of $w(t)$ in this case, we remind that each $e_{\alpha}$ must increase in time following the applied strain $\dot\gamma t$, with 
jumps when passing from one potential well to the next. We will consider that each $e_{\alpha}$ is thus a cumulative Poisson process, and that $w(t)$ is a sum with variable signs of many of these processes, so $w(t)$ turns out to be a random walk process.
Concerning the amplitude of the process $w(t)$, we notice that as this process is originated in the values of $e$ in different parts of the sample, the time scale must also be related to the average strain $\dot\gamma t$. This can be incorporated as a proportionality of the amplitude of $w(t)$ with $\sqrt{\dot\gamma}$.
Summarizing, breaking the self-consistency condition, the mean field equation leads to the truly one particle model
(now we also drop the ${\alpha}$ label, the equations apply to a generic site)
\begin{eqnarray}
\dot e&=& f(e)+\sigma+k(\dot\gamma t-e) +w(t) \label{stochastic1}
\\
\dot w&=&\nu\sqrt{\dot\gamma}\eta(t)
\label{stochastic2}
\end{eqnarray}
where $\eta(t)$ is an unitary variance delta correlated white noise: $\langle \eta(t)\rangle =0$, $\langle \eta(t)\eta(t')\rangle =\delta(t-t')$, and $\nu$ is a global amplitude of the fluctuating term.

The analysis of this {\em stochastically driven} PT model (Eqs. (\ref{stochastic1}) and (\ref{stochastic2})) is presented in Appendix B. There it is shown that the stochastic term produces a decrease of the critical stress, and -more importantly- a modification of the $\beta$ exponent. The value of $\beta$ without and with the stochastic term changes from $\beta=1$ to $\beta=2$ for parabolic potentials, and from $\beta=3/2$ to $\beta=5/2$ for smooth potentials (see Table 1).

\begin{figure}
\includegraphics[width=8cm,clip=true]{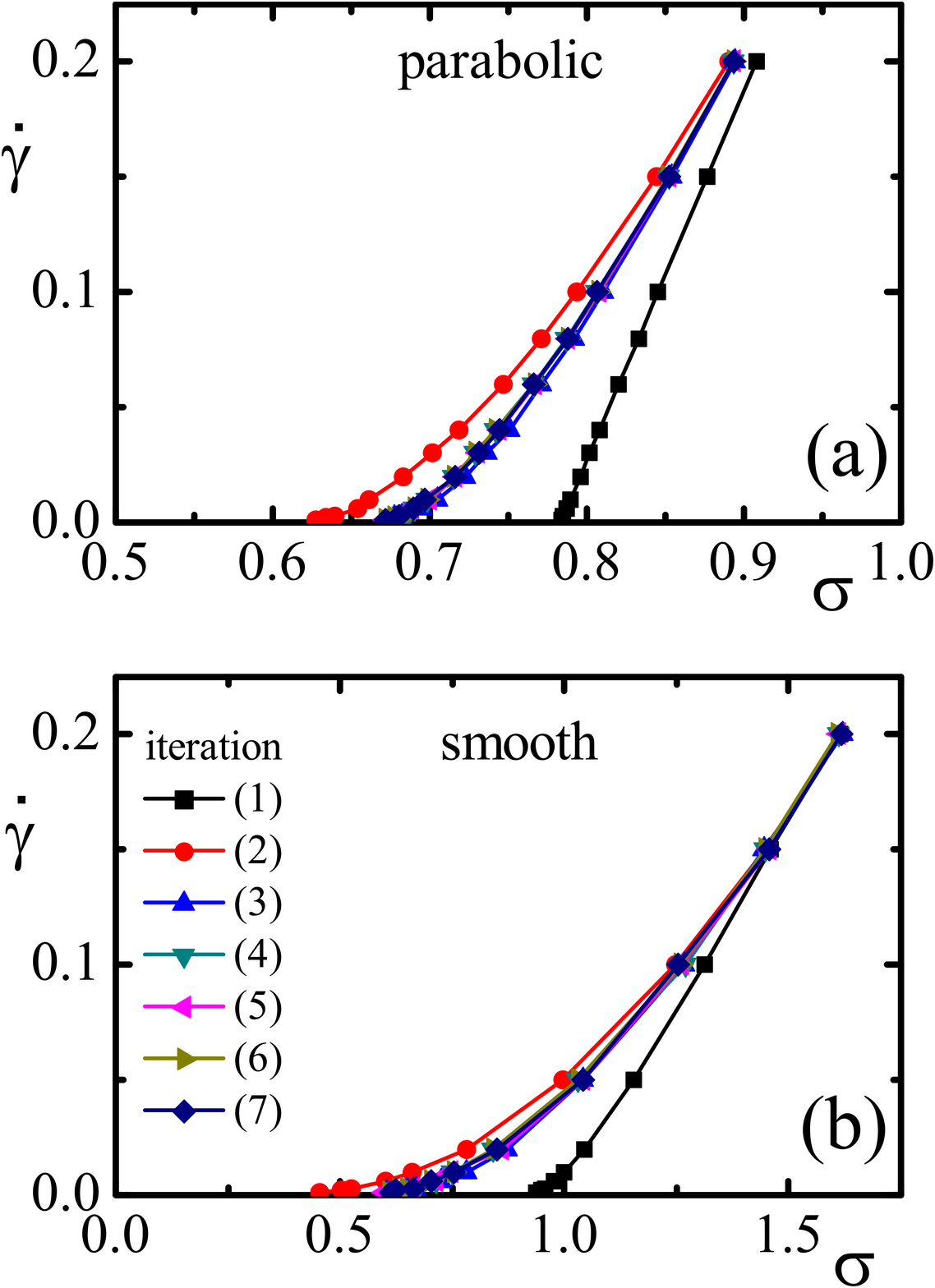}
\caption{Flow curves for the iterative implementation of the self-consistently driven PT model, for parabolic and smooth potentials. The first iteration gives the results of the standard PT model, and the second one corresponds to the stochastically driven PT model. After a few iterations the flow curves converge to a limiting curve with an intermediate value of the  $\beta$ exponent.
\label{sdpt1}
}
\end{figure}
\begin{figure}
\includegraphics[width=8cm,clip=true]{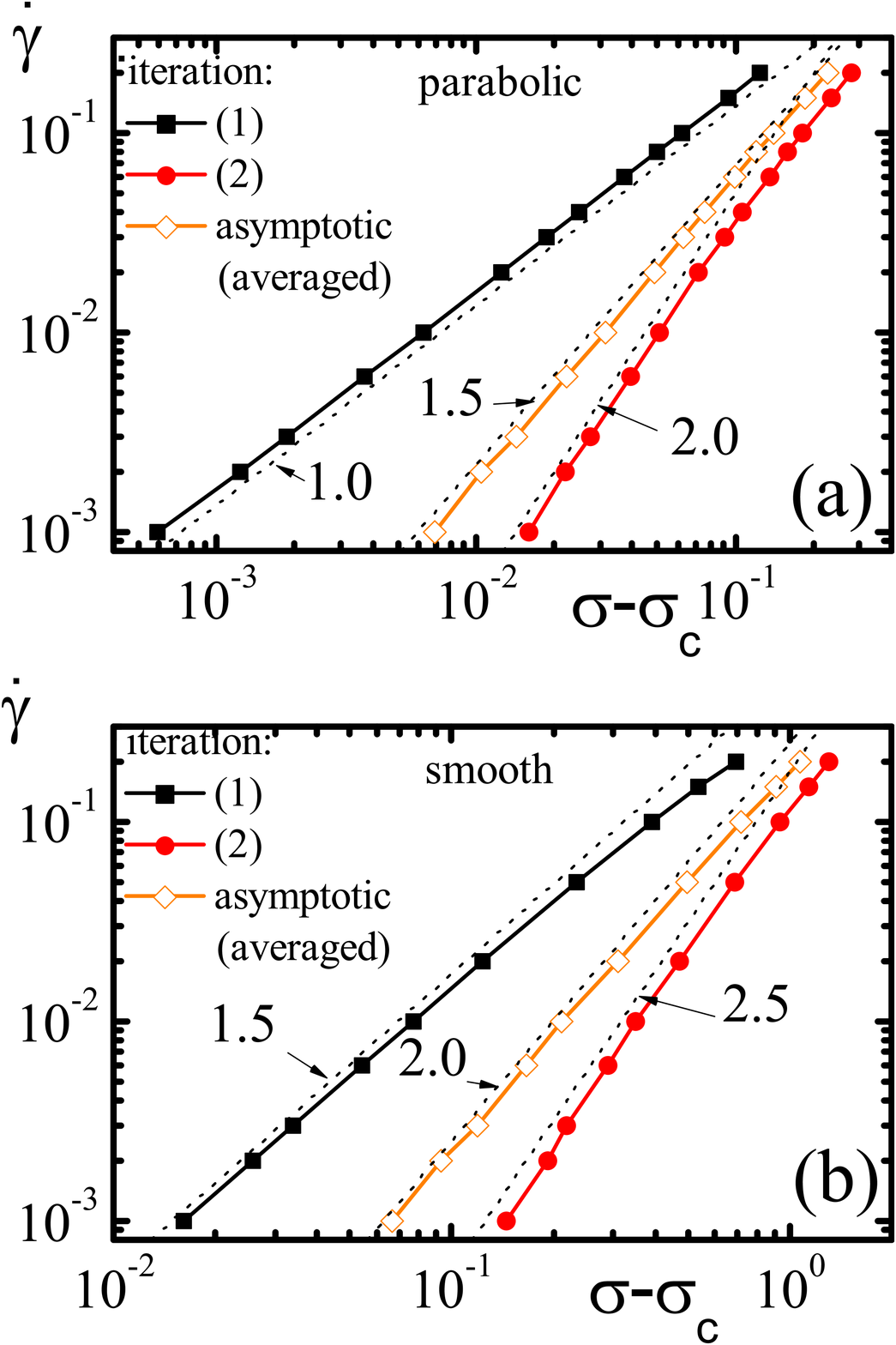}
\caption{The data from the previous figure corresponding to the first and second iteration, and the average from iterations ten to twenty, plotted as a power law around $\sigma=\sigma_c$. Dotted lines are guides to the eye, drawn with the indicated slope.
\label{sdpt2}
}
\end{figure}

We will now analyze the self consistently driven case and see that it generates intermediate values of $\beta$.
Unfortunately we have not been able to find an analytical solution for the self-consistently driven PT model, and had to rely on numerical simulations of Eqs. (\ref{sc3}) and (\ref{sc4}) in order to investigate the values of $\beta$ they provide.

We implemented a successive approximation scheme to solve (\ref{sc3}) and (\ref{sc4}) that goes as follows.
We take an ensemble of sites $e_{\alpha}$ and drive them with the uniform driving $e_0(t)\equiv \dot\gamma  t +\sigma/k$ alone. We call the results $e_{\alpha}^{(1)}(t)$. 
From them, a stochastic driving term is calculated as

\begin{equation}
w^{(1)}(t)=\sum_{\alpha} \lambda_{\alpha} e_{\alpha}^{(1)}(t)
\end{equation}
Then a fresh set of sites $e_{\alpha}$ are evolved under the driving $e_0(t)+w^{(1)}(t)$, obtaining new values $e_{\alpha}^{(2)}(t)$,
and the process is repeated. 



We present results of this iterative scheme for a system of $N=1000$ sites, with $k=.5$ (parabolic) and $k=.1$ (smooth), and values of $\lambda_{\alpha}$ taken from a normal distribution of zero mean and variance $1/N$ (parabolic) and $0.2/N$ (smooth).
In Fig. \ref{sdpt1} we show the values of $\sigma$ as a function of $\dot\gamma$ at the successive steps of the iteration procedure. The first step reproduces the behavior of the pure PT model. The second step 
corresponds to the stochastically driven case (see Appendix) as the driving comes from the composition of the driving of the uncorrelated PT particles of the first step.
Successive steps converge rapidly towards a flow curve with an intermediate value of the $\beta$ exponent.

In order to provide a numerical estimation of the self-consistent $\beta$ value, we average the results from steps (10) to (20) for which the data show already a good convergence, and fit them with expressions of the form $\dot\gamma\sim (\sigma-\sigma_c)^\beta$. The results are presented in Fig. \ref{sdpt2}. 

The obtained values of the exponent $\beta$ are clearly in between those of the normal PT model and those of the PT model with stochastic driving, indicating first of all that the  self consistent driving
is a non-trivial ingredient that affects the behavior of the system. The numerical values are estimated as $\beta=1.5\pm 0.2$ in the parabolic case, and $\beta=2.0\pm 0.2$ in the smooth case.
These values, obtained in a mean-field-model, and taking into account the numerical uncertainties,  strongly suggest the possibility that the exact values are 3/2, and 2. Unfortunately, at present we have no proof of this conjecture.
Moreover, we also note that the values found with the self-consistent PT model are  compatible with those obtained in the simulation of the full model (Fig. \ref{f3}, and Table I). The question then remains if this indicates just a proximity of the values, or if the values of $\beta$ in the full and mean field models are exactly the same.

\begin{table}
\begin{center}
\begin{tabular}{ |c|c|c| } 
 \hline
flow exponent $\beta$  & parabolic & smooth \\ 
 &  potential &  potential \\ 
 \hline
full simulation & $\sim$ 1.5 & $\sim$ 2.0 \\ 
 PT model& 1 & 3/2 \\ 
 stochastically driven PT & 2 & 5/2 \\ 
self-consistent PT  & $\sim$ 1.5 & $\sim$ 2.0 \\ 
 \hline
\end{tabular}
\end{center}
\caption{Summary of values of the flow exponent $\beta$ found in this work, for the two kinds of potentials analyzed. Approximate results from numerical simulations are preceded by ``$\sim$". Other values are exact.}
\label{table:1}
\end{table}

Beyond the dependence of the $\beta$ values on the particular approximation scheme used, the results in Table I strongly support the existence of systematic differences between the values obtained using smooth or parabolic potentials. We argue on the reason of this difference in the next Section.

\section {Comparison with the depinning case}

Depinning models with local elastic interactions are typically described by equations like

\begin{equation}
\frac{dx_i}{dt}=f_i(x_i) +k\left (n^{-1}\sum_{j=1}^n x_{j}-x_i\right)+\sigma
\label{dep}
\end{equation}
where $x_i$ are the elastic deformations, the sum runs over the $n$ neighbors to site $i$, and $f_i$ is the local pinning force.
The ``fully connected" version of this model (in which any site interacts equally with any other of the $N$ sites in the system) leads to the ``mean-field-like" equation 
\begin{equation}
\frac{dx_i}{dt}=f(x_i) +k(\overline x-x_i)+\sigma
\label{dep_mf}
\end{equation}
where $\overline x=\sum x_i/N$. This equation 
has the form of a PT model, and so it provides different values of $\beta$  for parabolic and smooth pinning potentials (namely $\beta=1$ and 3/2, respectively). 
This was already pointed out by Fisher is his seminal studies of depinning of charge density waves\cite{f1,f2}.
Yet, for depinning with short range elastic interactions (Eq. (\ref{dep}))
the value of $\beta$ is known to be independent of the kind of potential used. In particular, $\beta=1$ represents the correct mean field exponent, for both kinds of potentials. The reason is very subtle, and it has to do with the analysis of the model upon renormalization. It is demonstrated using functional renormalization group theory \cite{frg1,frg2,frg3} that even if local smooth potentials are used, the effective pinning potential becomes singularly correlated upon renormalization, and the renormalized potential develops cusps that make the result independent of the detailed form of the starting potential.

On the contrary, for the case of yielding the results of the present numerical simulations
show persisting differences between the two kinds of potentials used, in particular the values of $\beta$ differ for smooth and parabolic potentials. 
Our interpretation of this behavior is related to the existence in the effective scalar equation of the model (Eq. (\ref{tomlinson})) of the infinite range term proportional to $k$. Note that this term appears as a consequence of the elasticity of the system, and is not originated in any kind of mean field approximation. 
This kind of terms have been obtained in other contexts, for instance in \cite{nelson,pepe,saizer}.
The dependence of the value of $\beta$ on the smooth/parabolic form of the potential in Eq. (\ref{tomlinson}), is exactly the same dependence that Eq. (\ref{dep_mf}) displays, with the additional ingredient given by the Eshelby elastic interaction in Eq. (\ref{tomlinson}).
This term, having also a long range effect ($\sim r^{-2}$) seems to be capable of modifying the values of $\beta$ that 
would appear if it was absent.  
Yet, it does not erase the differences between the two kinds of potentials.

\section{Conclusions}

In this work we have studied a mesoscopic model for the yielding transition of a two-dimensional amorphous material under an externally applied deviatoric deformation. The model incorporates in a realistic way the elastic deformations of the material, and in particular the way in which these deformations at some part of the sample affect other regions of the material. Plastic deformation is accounted for by introducing local disordered ``plastic potentials" for the deformation, allowing for each piece of the system to jump among different minima of these potentials, representing different structural configuration with different strain. 

We have observed that this model displays a well defined yielding point, i.e., a minimum shear stress $\sigma_c$ has to be applied in order for the system to deform at a constant strain rate $\dot\gamma$, no matter how small. Around the yielding point, the strain rate and the stress are power law related: $\dot\gamma\sim(\sigma-\sigma_c)^\beta$.
The main result we have obtained is that the value of $\beta$ depends on the form of the plastic potential that is used. For smooth potentials we find $\beta\simeq 2.0$, whereas for potentials formed by a concatenation of parabolic pieces, a value $\beta\simeq 1.5$ is obtained. 
These results indicate that there is more than one universality class associated to yielding, contrary to the well established result of a single universality class for the related problem of elastic depinning in low dimensions. 

In addition, we have derived a simplified scalar version of the model that has the form of a set of 
Prandtl-Tomlinson particles, coupled by a quadrupolar Eshelby interaction. We have done a mean field approximation on the quadrupolar term, finding values of $\beta$ compatible with those of the full simulation, and in particular a persistent difference between the values for smooth and parabolic potentials.
We interpret this persistent difference as originated in the global coupling of the Prandtl-Tomlinson particles to the mean global coordinate. This interaction is a direct consequence of the material elasticity and does not emerge from any kind of approximation.

Although we have obtained differences in other exponents for the smooth and parabolic cases, the numerical quality of those results is not satisfactory at present. Further studies are thus necessary to elucidate if this problem can in fact be consistently described as possessing two different universality classes with two different sets of critical exponents.

\section{Acknowledgments}

I thank Ezequiel Ferrero for helpful discussions, and Craig Maloney for many insightful comments on a first version of the manuscript. 

\appendix

\section{Details on the form of the plastic potentials}

\begin{figure}
\includegraphics[width=7cm,clip=true]{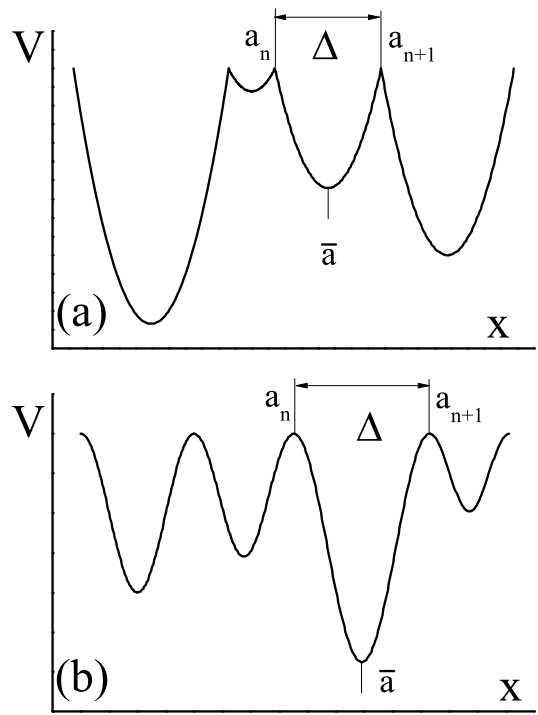}
\caption{Typical plastic potentials that are generated for the parabolic case (a) and the smooth case (b).
Note that the curvature of the potentials at all minima is the same
\label{potencial}
}
\end{figure}

Here we provide details on the way in which the plastic potentials (sketched in Fig. \ref{f1}) are actually constructed. For each site $i$ in the system a potential $V_i(x_i)$ is constructed, that has a stochastic ingredient. For different sites, the stochastic component is chosen in an uncorrelated way. A generic potential $V(x)$ is constructed piecewise, by dividing the $x$ axis in segments through a set of values $a_n$
(see Fig. \ref{potencial}). In each interval $a_n$-$a_{n+1}$ (defining $\overline a\equiv (a_{n+1}+a_{n})/2$, and $\Delta \equiv a_{n+1}-a_{n}$) the potential is defined as
\begin{equation}
V(x)=\frac 12 \left [(x-\overline a)^2-\Delta^2\right ]
\end{equation}
in the parabolic case, and
\begin{equation}
V(x)=-\left (\frac{\Delta}{2\pi}\right )^2\left [1+\cos\left(\frac {2\pi (x-\overline a)}\Delta\right)\right ]
\end{equation}
in the smooth case. Note that even in the smooth case the potential is not analytic, but it has a continuous second derivative, which is enough for our purposes. Also, the curvature of the potential in all minima is the same, and this is chosen to have an isotropic elastic medium in the harmonic approximation. 
The separation $\Delta$ between $a_n$ and $a_{n+1}$ is stochastically chosen from a flat distribution between $\Delta_{min}=2$ and $\Delta_{max}=4$.

\section{The stochastically driven Prandtl-Tomlinson model}

In this appendix we make a dimensional analysis of a generalized PT model, in which in addition to the deterministic driving at a constant velocity, there is also a stochastic term with the characteristics of a random walk, as represented by Eqs. (\ref{stochastic1}),(\ref{stochastic2}). For the present purposes, these equations can be conveniently written as

\begin{eqnarray}
\dot e=f(e)+k(w(t)-e)\label{zz1}\\
\dot w=\dot\gamma +\nu\sqrt{ \dot\gamma}\eta(t)\label{zz2}
\end{eqnarray}
Note that the deterministic part of the driving was included in the equation for $\dot w$.

In the case $\nu=0$ the problem reduces to the usual PT model. This model displays a non-zero critical force $\sigma_c$ (at vanishingly small $\dot\gamma$) when the pinning force $f(x)$ is sufficiently strong. For finite $\dot\gamma$ the friction force increases according to $\sigma-\sigma_c \sim \dot\gamma^{1/\beta}$. We recall the arguments leading to the determination of the value of $\beta$, taking advantage of a dimensional analysis.
The time scale of the dynamics at very small $\dot\gamma$ is dominated by the surpassing of the energy barriers of the pinning energy, namely by the maxima of $f(e)$. 
Around one of these maxima (assumed to occur at $e=0$) we can write $f(e)\simeq D|e|^\alpha$.
For smooth pinning potentials $\alpha=2$, whereas for a concatenation of parabolas $\alpha=1$. We keep a general exponent $\alpha$ for the analysis.

For a narrow interval of the variable $e$ around zero the last term in Eq. \ref{zz1} can be neglected, and equation of motion for $e$ can be written as

\begin{equation}
\dot e= D|e|^\alpha +kw(t)= D|e|^\alpha +k\dot\gamma t
\label{4}
\end{equation}
where time is set as zero at the moment in which the driving is able to overcome the energy barrier.
 For $\dot\gamma \to 0$,  $e$ reaches the top of the barrier (i.e, $e=0$) at $t=0$. For finite $\dot\gamma$ there will be a delay in reaching the $e=0$ point. This delay is the main responsible of the increase of the friction force with $\dot\gamma$. In order to obtain the dependence of the delay with $\dot \gamma$ we can rescale Eq. \ref{4} in order to eliminate $\dot\gamma$. Defining
\begin{eqnarray}
\hat e= (k\dot \gamma)^{\frac{-1}{2\alpha-1}}D^{\frac 2{2\alpha-1}}e\\
\hat t=(k\dot \gamma)^{\frac{\alpha-1}{2\alpha-1}}D^{\frac 1{2\alpha-1}}t
\end{eqnarray}
Eq. \ref{4} can be written as
\begin{equation}
\dot {\hat e}= |\hat e|^\alpha +\hat t
\end{equation}
In this form it is clear that there will be a single solution $\hat e(\hat t)$ for all values of $\dot \gamma$.
The time at which $e$ reaches the instability value $0$ will correspond to a single value $\hat \tau$ of $\hat t$. In the original units this will give the time values as $\tau(\dot\gamma)\sim \dot\gamma ^{\frac{1-\alpha}{2\alpha-1}}$. By this time, the value of the driving $w(t)$ has reached a value $w(\tau)=\dot \gamma \tau(\dot\gamma)\sim \dot\gamma^{\frac{\alpha}{2\alpha-1}}$, and this represents an increase of the friction force compared to the $\dot \gamma=0$ case of $\sigma-\sigma_c\sim \dot\gamma^{\frac{\alpha}{2\alpha-1}}$, i.e. $\beta=2-1/\alpha$. We get $\beta=3/2$ for
$\alpha=2$ (the standard case of smooth potentials) and $\beta=1$ for $\alpha=1$ (for a potential that is constructed as a concatenation of parabolas). Both these values of $\beta$ are well known in the context of the PT model.

Now in the presence of a stochastic component of the driving, the equivalent to Eq. \ref{4} reads
\begin{equation}
\dot e= D|e|^\alpha +kw(t)
\label{ee}
\end{equation}
with
\begin{equation}
\dot w(t)= \dot\gamma+ \nu\sqrt{\dot \gamma}\eta(t)
\end{equation}
where $\eta(t)$ is an uncorrelated noise, i.e, $\langle\eta(t)\rangle=0$, $\langle\eta(t)\eta(t')\rangle=\delta(t-t')$. The dominant contribution to calculate the flow exponent $\beta$ comes in this case from the fluctuating term in the driving, and searching for this contribution we can neglect for the moment the linear part of the driving. In this way, we can analyze the case in which 
\begin{equation}
\dot w(t)= \nu\sqrt{\dot \gamma}\eta(t)
\label{5}
\end{equation}

Proceeding as before, we rescale $e$ and $t$ in order to eliminate $\dot\gamma$ from \ref{ee}-\ref{5}.
Defining
\begin{eqnarray}
\hat e=(k^2\nu^2{\dot \gamma})^{\frac{-1}{3\alpha-1}}D^{\frac{3}{3\alpha-1}}e\\
\hat t=(k^2\nu^2{\dot \gamma})^{\frac{\alpha-1}{3\alpha-1}}D^{\frac{2}{3\alpha-1}}t
\end{eqnarray}
Eqs. \ref{ee}-\ref{5} read

\begin{eqnarray}
\dot {\hat e}= |\hat e|^\alpha +\hat w(\hat t)\\
\dot {\hat w(\hat t)}= \eta(\hat t)
\end{eqnarray}
and this shows there will be a single value $\hat \tau$ of the delay time for any $\dot\gamma$. In the original variables we obtain the dependence of the delay time with $\dot\gamma$ as $\tau(\dot\gamma)\sim \dot\gamma^{\frac{1-\alpha}{3\alpha-1}}$. By this time, the stochastic driving attains a value $\sim \sqrt{\dot\gamma \tau(\dot\gamma)}\sim \dot\gamma^{\frac{\alpha}{3\alpha-1}}$, from which we obtain in this case $\beta=3-1/\alpha$, which is 2 for parabolic potentials, and 5/2 for smooth potentials.

\begin{figure}
\includegraphics[width=9cm,clip=true]{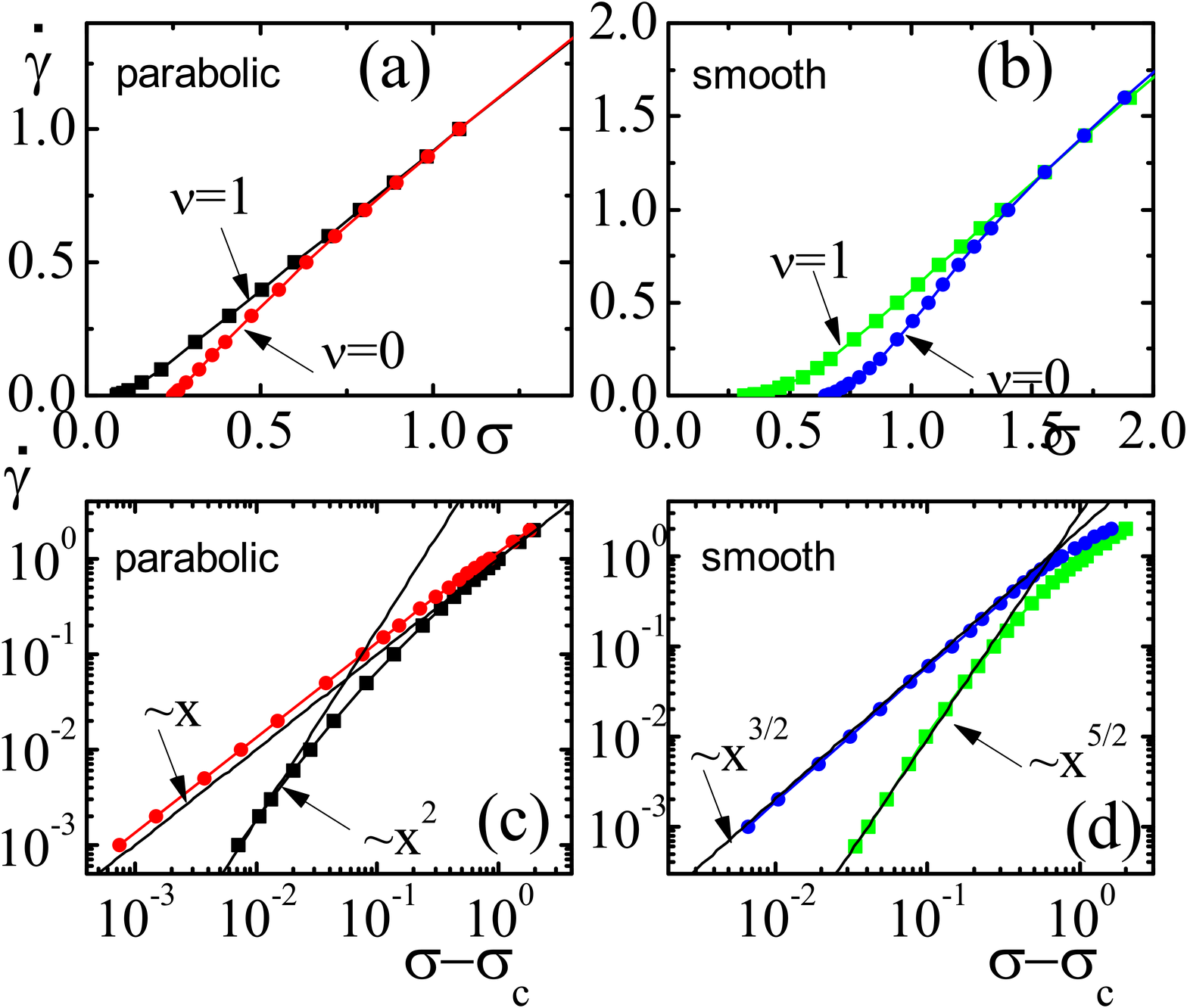}
\caption{Results for the standard ($\nu=0$) and stochastically driven ($\nu=1$) PT model (Eqs. \ref{zz1} and \ref{zz2}), for the case of parabolic and smooth potentials. Panel (a) and (b) are the results in linear scale, whereas (c) and (d) are in logarithmic scale, with $\sigma$ shifted in each case by the numerically determined $\sigma_c$. The asymptotic forms (dotted lines) display the exponents predicted by the analytical treatment. The numerical data tend to match the analytical behavior in the small $\dot\gamma$ limit.
\label{1pla}
}
\end{figure}

To our knowledge, the PT model in the presence of this kind of stochastic driving has not been analyzed before. It seems thus appropriate to present results of direct numerical simulations in order to
verify the previous analytical estimations and to see how the full curve $\sigma(\dot\gamma)$ looks like.
We simulate Eqs. \ref{zz1} and \ref{zz2}, with the particular choice $f(e)=\sin(2\pi e)$ for the smooth potential case, and $f(e)=-(2e-[2e])/2$ (where $[x]$ is the nearest integer to $x$) for the parabolic potential case.
Simulations are straightforward, and are done with a first order Euler method, with time step $10^{-3}$ and  $k=1$.
Results are contained in Fig. \ref{1pla}. They show that the presence of the stochastic term reduces the value of $\sigma_c$, and -most importantly- changes the value of $\beta$. The values $\beta=2$, and $\beta=5/2$ for parabolic and smooth potentials respectively are accurately obtained in the simulations in the limit of very small $\dot\gamma$.

\begin{thebibliography}{37}
\bibitem{e2e3}The choice between applying the external driving on $e_2$ or $e_3$ (and the corresponding plastic response) is irrelevant in the ideal case of continuous, infinite systems since the two choices are related by the rotation of 45 degrees of an isotropic system. However, when putting the model on a finite square mesh, we should be careful about locking effects between the mesh and the driving. The choice of $e_2$ for driving is expected to reduce these locking effects since the easy slip directions in this case do not align (actually, they are at 45 degrees) with the coordinate axis. Additional runs using $e_3$ as the driving show in particular that the peculiarities we find about the values of $\beta$ persist, i.e., they are not an artifact originated in the numerical mesh. The values of the yield stress do change however.
\bibitem{nota_lambda}This can be understood in the following way. For sufficiently large $N$, there will always be sites that are almost equivalent to each other (in the sense that the have almost the same plastic potential). In Eq. (\ref{sc4}) all that matters is thus the sum of the $\lambda_\alpha$ of all these equivalent sites. This sum will have the form of a normal variable in the $N\to\infty$ limit.
\end{thebibliography}
\end{document}